\documentclass{amsart}[12pt]

\usepackage{amssymb}
\usepackage{graphicx}
\usepackage{caption2}
\usepackage{amsfonts}
\usepackage{cite}
\usepackage{lineno}
\usepackage{color}
\usepackage{multirow}
\usepackage{subfigure}
\usepackage{bm}

\newcommand{\be}{\begin{equation}}
\newcommand{\ee}{\end{equation}}
\newcommand{\bea}{\begin{eqnarray}}
\newcommand{\eea}{\end{eqnarray}}
\newcommand{\bef}{\begin{figure}}
\newcommand{\ef}{\end{figure}}
\newcommand{\bt}{\begin{tabular}}
\newcommand{\et}{\end{tabular}}
\newcommand{\bno}{\begin{enumerate}}
\newcommand{\eno}{\end{enumerate}}

\allowdisplaybreaks[4]

\def\3{\ss}

\catcode`\"=\active
\def"{\accent'177}

\begin{document}

\begin{center}

{\Large\bf  Discovery of 10,059 new three-dimensional periodic orbits \\ of general three-body problem} 

\vspace{0.3cm}

Xiaoming Li$^{1}$ and Shijun Liao$^{2, 3 *}$ \\
\vspace{0.25cm}
$^{1}$School of Mechanics and Construction Engineering, Jinan University, Guangzhou 510632,
China \\
$^{2}$School of Ocean and Civil Engineering, Shanghai Jiao Tong University, Shanghai 200240, China\\
$^{3}$School of Physics and Astronomy, Shanghai Jiao Tong University, Shanghai 200240, China\\

$^*$Corresponding author: sjliao@sjtu.edu.cn

\end{center}

\hspace{-0.75cm}{\bf Abstract} 
{  \em
A very few three-dimensional (3D) periodic orbits of general three-body problem (with three finite masses)  have been discovered since  Newton mentioned it in 1680s.   
Using a high-accuracy numerical strategy  we discovered 10,059 three-dimensional  periodic orbits of the three-body problem in the cases of $m_{1}=m_{2}=1$ and $m_{3}=0.1n$  where $1\leq n\leq 20$ is an integer, among which  1,996 (about 20\%)  are linearly stable.  Note that our approach is valid for arbitrary mass  $m_{3}$  so that in theory we can gain an arbitrarily large amount of 3D periodic orbits of the three-body problem.
In the case of three equal masses,  we discovered twenty-one 3D ``choerographical''  periodic orbits whose three bodies move periodically in a single closed orbit.  It is very interesting that,  in the case of two equal masses, we discovered  273 three-dimensional periodic orbits with the two bodies ($m_{1}=m_{2}=1$) moving along a single closed orbit and the third ($m_{3}\neq 1$)  along a different one: we name them ``piano-trio'' orbits, like a trio for two violins and one piano.  To the best of our knowledge, all of these 3D periodic  orbits have never been reported,  indicating the novelty of this work.  The large amount of these new 3D periodic orbits are helpful for us to have  better understandings about chaotic properties of the famous three-body problem, which ``are, so to say, the only opening through which we can try to penetrate in a place which, up to now, was supposed to be inaccessible'', as pointed out by Poincar{\'e}, the founder of chaos theory. 
}

\hspace{-0.75cm}{\bf Keywords} {general three-body problem, three-dimensional periodic orbits }




\vspace{0.5cm}

\section{Introduction}

The famous three-body problem can be traced back to Newton \cite{Newton1687} in 1680s,  which focuses on the motion of three bodies with finite masses that attract to each other by gravitational force under Newton's second law.  It attracted many famous mathematicians and physicists such as Euler~\cite{Euler1767} and Lagrange~\cite{Lagrange1772} who found a very few of the earliest planar  periodic orbits of the  three-body problem.  The three-body problem is  three-dimensional (3D) in essence, but in the three hundred years after Newton mentioned it, a very few periodic orbits of the general three-body problem were found almost in the two-dimensional (2D) case  that is certainly much simpler than the three-dimensional one.  It was  Poincar{\'e}~\cite{Poincare1890}  who  revealed its mathematical complexity and difficulty in 1890:  the first integrals for the motion of the three-body system do not exist so that its closed-form solution is impossible in general, and besides its trajectories have the sensitivity dependence on initial conditions that laid the foundation of modern chaos theory.  
This well explains why in the 300 years only three families of planar periodic orbits of three-body system were found by Euler~\cite{Euler1767}  in 1767 and Lagrange~\cite{Lagrange1772}  in 1772, until 1970s when the Broucke-Hadjidemetriou-Henon family of planar periodic orbits~\cite{Broucke1973, Hadjidemetriou1975, Henon1976} were discovered  by means of computer.  In 1993 the so-called Figure-8  planar periodic orbit was numerically discovered by Moore~\cite{Moore1993} and subsequently rediscovered and rigorously proven by Chenciner and Montgomery~\cite{Chenciner2000}, which is called ``choreographic'' orbit:  the three bodies move periodically in a single closed orbit.  In 2013, \v{S}uvakov and Dmitra\v{s}inovi\'{c}~\cite{Suvakov2013}  made a breakthrough to numerically discover 13 new distinct planar periodic orbits by means of modern computer using Eulerian collinear initial configuration that has since been widely adopted to search for planar  periodic orbits~\cite{Iasko2014,Hudomal2015, Suvakov2016, Rose2016}. 

Why is it so difficult to find periodic orbits of three-body system even in 2013 when there exist supercomputers with very high performance?   It was Lorenz~\cite{Lorenz2006Tellus} who revealed the reason in 2006:  despite the sensitivity dependence on initial condition~\cite{Poincare1890},  a  chaotic system has also the sensitivity dependence on numerical algorithm so that chaotic trajectory is quickly polluted by numerical noise.  Thus,  since three-body problem is essentially chaotic according to Poincar{\'e}~\cite{Poincare1890},  its trajectory should be sensitive to numerical noises that are unavoidable for traditional numerical algorithms  such as Runge-Kutta's method (in double precision) that is widely used to gain trajectories of three-body system.  In 2009 Liao~\cite{Liao2009} proposed  the so-called ``clean numerical method'' (CNS)~\cite{Liao2014-SciChina, Li2018clean, Hu2020, qin2020influence, Liao2023} whose numerical noises can be reduced to such a low level that accurate/convergent chaotic trajectories can be gained in a finite but long enough time  interval $t\in[0,T_{c}]$, where the so-called ``predictable critical time'' $T_{c}$ is determined by the level of background numerical noise.  We emphasize here that accurate chaotic trajectory is very important for three-body problem, since it is the key reason why the number of discovered new families of planar periodic orbits of general three-body problem increases to several orders of magnitude:  in 2017 Li and Liao \cite{Li2017} discovered more than 600 new families of planar periodic orbits of three-body problem with three equal masses by means of computer using a numerical strategy based on CNS, then in 2018  Li, Jing and Liao~\cite{Li2018}  further discovered 1,223 new families of planar periodic orbits of three-body problem with two equal masses in a similar way, 
in 2021 Li~et~al.~\cite{Li2021} successfully obtained 135,445 new planar periodic orbits (including 13,315 stable ones) with arbitrarily unequal masses, and in 2022 Liao et al.~\cite{Liao2022}  proposed an effective roadmap to numerically gain planar periodic orbits of three-body systems with arbitrary masses by means of CNS and machine learning.  
In 2024 Hristov~et~al.~\cite{Hristov2024a} found 24,582  equal-mass periodic orbits of free-fall  three-body problem by means of high-accuracy algorithm.  
Finally, there are no any obstacles for us to gain masses of new {\em planar} periodic orbits of three-body system with arbitrary masses by means of high-accuracy numerical algorithms such as CNS.            

However,  compared to the planar periodic orbits mentioned above, the three-dimensional (3D) periodic orbits of  three-body problem are far limited.  Two types of periodic orbits were mainly studied.  The first is the spatial isosceles three-body problem, in which two bodies have equal masses and the third body oscillates along a straight line, while the three bodies always form an isosceles triangle.  In the so-called restricted three-body problem that contains one massless body, this configuration is known as the Sitnikov problem \cite{Sitnikov1961}, where two massive bodies move along an elliptical orbit in a plane, and the massless body oscillates in a straight line perpendicular to the plane.  
This class of 3D periodic orbits has been extended to the general three-body problem, where all three bodies have finite masses \cite{Corbera2004, Yan2015, Perdomo2017}.  The second types were gained using the continuation methods to generate 3D periodic orbits from periodic orbits of restricted three-body problem \cite{Katopodis1979, Markellos1980}.  It is a great pity that a very few  3D periodic orbits of general three-body problem have been found, to the best of our knowledge.  Thus, it has great scientific meanings to discover a masses of 3D periodic orbits of general three-body problem with finite masses.

\section{Initial configuration and numerical strategy} 

\begin{figure}[tb]
\centering
  \includegraphics[width=8cm]{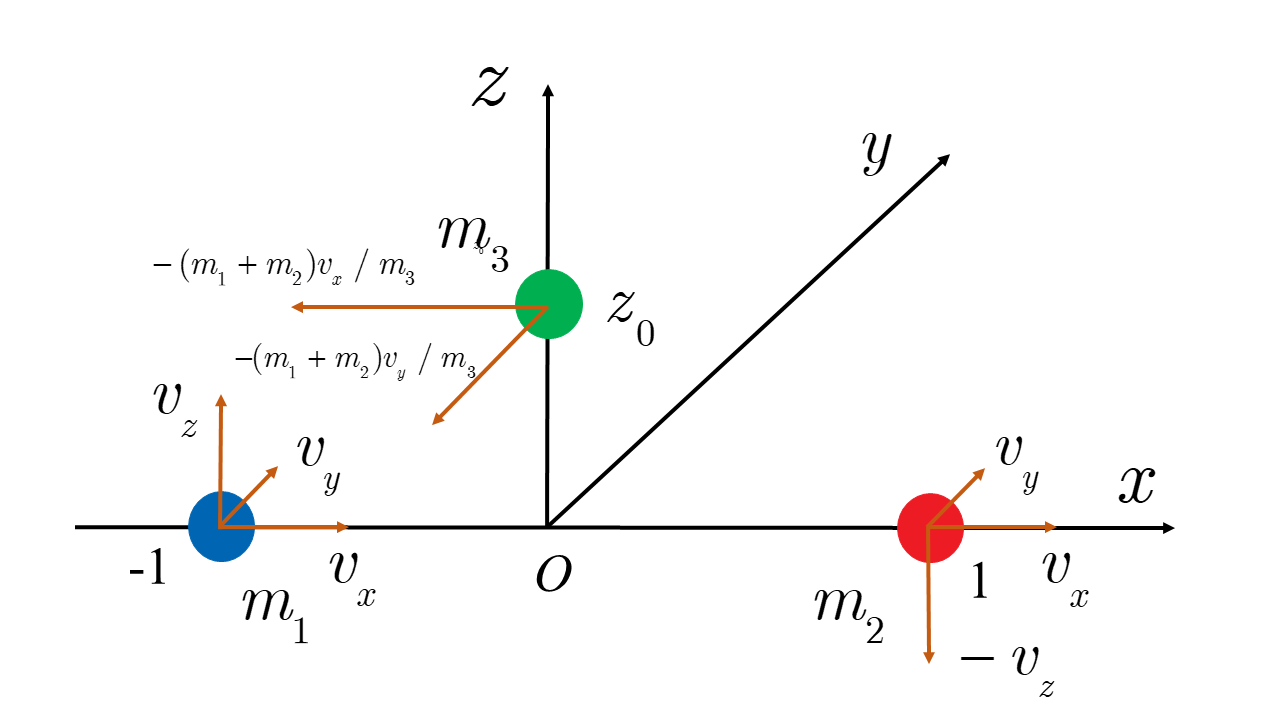}
  \caption{The initial configuration of the 3D three-body system. The initial positions of the three bodies are located at $\bm{r}_1 = (-1, 0, 0)$, $\bm{r}_2=(1, 0, 0)$ and $\bm{r}_1 = (0, 0, z_{0})$ with the corresponding initial velocities $\bm{v}_1 = (v_x, v_y, v_z)$, $\bm{v}_2 = (v_x, v_y, -v_z)$ and $\bm{v}_3 = (-2 v_x/m_3, -2 v_y/m_3, 0)$.}\label{fig:IC}
\end{figure}

The  motions of the general three-body system (with finite masses) are governed by the dimensionless equations  
\begin{equation}
\frac{\mathrm{d}^2\bm{r}_{i}}{\mathrm{d}t^2}=-\sum_{j=1,j\neq i}^{3}  m_{j}\frac{(\bm{r}_{i}-\bm{r}_{j})}{| \bm{r}_i-\bm{r}_j |^{3}}, \quad i=1,2,3, \label{ODE}
\end{equation}
where $t$ denotes the time,  $\bm{r}_i$ and $m_i$ are the position vector and mass of the $i$-th body, respectively. 
Let $\bm{X}(t)=(\bm{r}_1(t), \bm{r}_2(t), \bm{r}_3(t), \dot{\bm{r}}_1(t), \dot{\bm{r}}_2(t), \dot{\bm{r}}_3(t))$  denote the state space of the three-body system at the time $t$, where $\dot{\bm{r}}_i(t)$ represents the velocity of the $i$-th body.
To evaluate the proximity of the system to its initial state $\bm{X}(0)$, we define the so-called  ``return proximity function'' 
\begin{eqnarray}
\delta(t) &=& |\bm{X}(t)-\bm{X}(0)|.\label{return-function}
\end{eqnarray}
 A periodic orbit is determined by $\delta(T)=0$, where $T$ is its period.  

As shown in Fig.~\ref{fig:IC},  we use an initial configuration for the 3D three-body system in this paper.  
The initial positions of the three bodies are located at $\bm{r}_1 = (-1, 0, 0)$, $\bm{r}_2=(1, 0, 0)$ and $\bm{r}_3 = (0, 0, z_0)$,  the  initial velocities of Body-1 and Body-2 are $\bm{v}_1 = (v_x, v_y, v_z)$ and $\bm{v}_2 = (v_x, v_y, -v_z)$, respectively, where $z_{0}, v_x, v_y, v_z$ are unknown.   
For the sake of simplicity,  let us consider  the case of zero momentum, i.e.  $m_1\bm{v}_1+m_2\bm{v}_2+m_3\bm{v}_3=0$, which gives the initial velocity   $\bm{v}_3 = (-(m_1+m_2)v_x/m_3, -(m_1+m_2)v_y/m_3, 0)$ of Body-3.  Note that in the above-mentioned configuration the initial condition of the 3D three-body system is determined  by the four unknown  variables   $v_x$, $v_y$, $v_z$, and $z_0$.

Our numerical strategy is briefly described as follows.  
First of all, we use the grid search method to search for possible candidates of 3D periodic orbits of the three-body system in the case of  $v_z$ = 0 
 within $z_0 \in (0, 1] $, $v_x \in (0, 1]$ and $v_y \in (0, 1]$:  1000 isometric points are used in each dimension, i.e.  for a chosen $m_{3}$ totally $1000 \times 1000 \times 1000 = 10^{9}$ cases are considered as initial condition to integrate the governing equations (\ref{ODE}) using the DOP853 solver \cite{Hairer1993}.   One among these initial conditions is regarded as a candidate of 3D periodic orbit if its corresponding return proximity function  (\ref{return-function}) is less than $10^{-1}$ under a value of period $T$.    
Secondly, the values of $(z_0, v_x, v_y, v_z)$ and $T$ of each candidate are constantly  modified/corrected  by means of the numerical approach  (similar to \cite{Li2018}) based on  the Newton-Raphson method \cite{Abad2011} and Clean Numerical Simulation (CNS) \cite{Liao2009, Liao2014-SciChina, Li2018clean, Hu2020, qin2020influence, Liao2023} :  a 3D periodic orbit is discovered when  the return proximity function $\delta(T)$ defined by  (\ref{return-function}) is less than $10^{-10}$.   The linear stability of periodic orbits is evaluated by Floquet theory \cite{Guckenheimer1983}.  To compute the monodromy matrix, we integrate the variational equations along periodic orbit \cite{Richter1993}. When all eigenvalues of the monodromy matrix lie on the unit circle, the periodic orbit is linearly stable. 
 
\section{Discovery of 10,059 three-dimensional periodic orbits}

\begin{figure}[tb]
\centering
  \includegraphics[width=5cm]{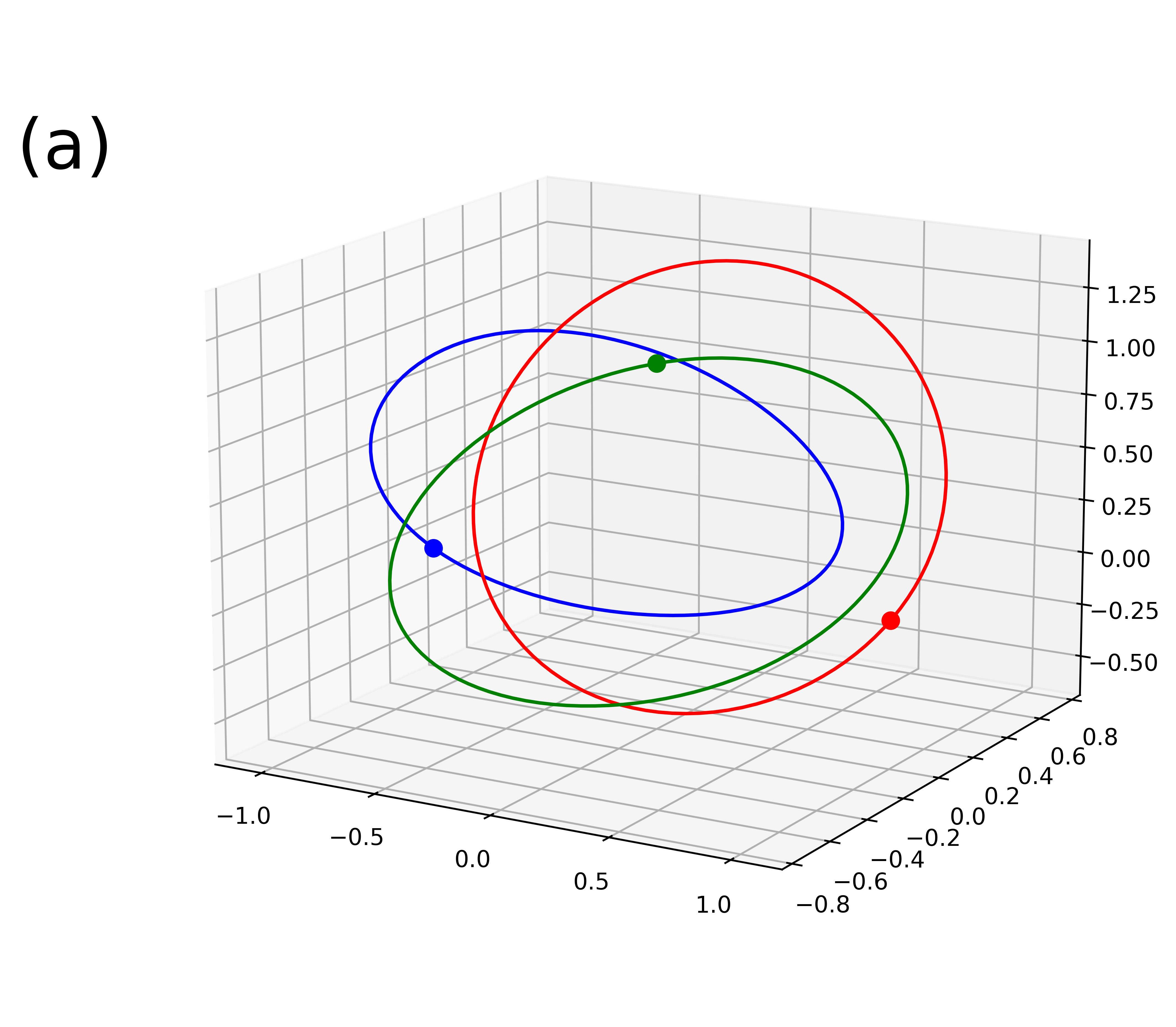}
  \includegraphics[width=5cm]{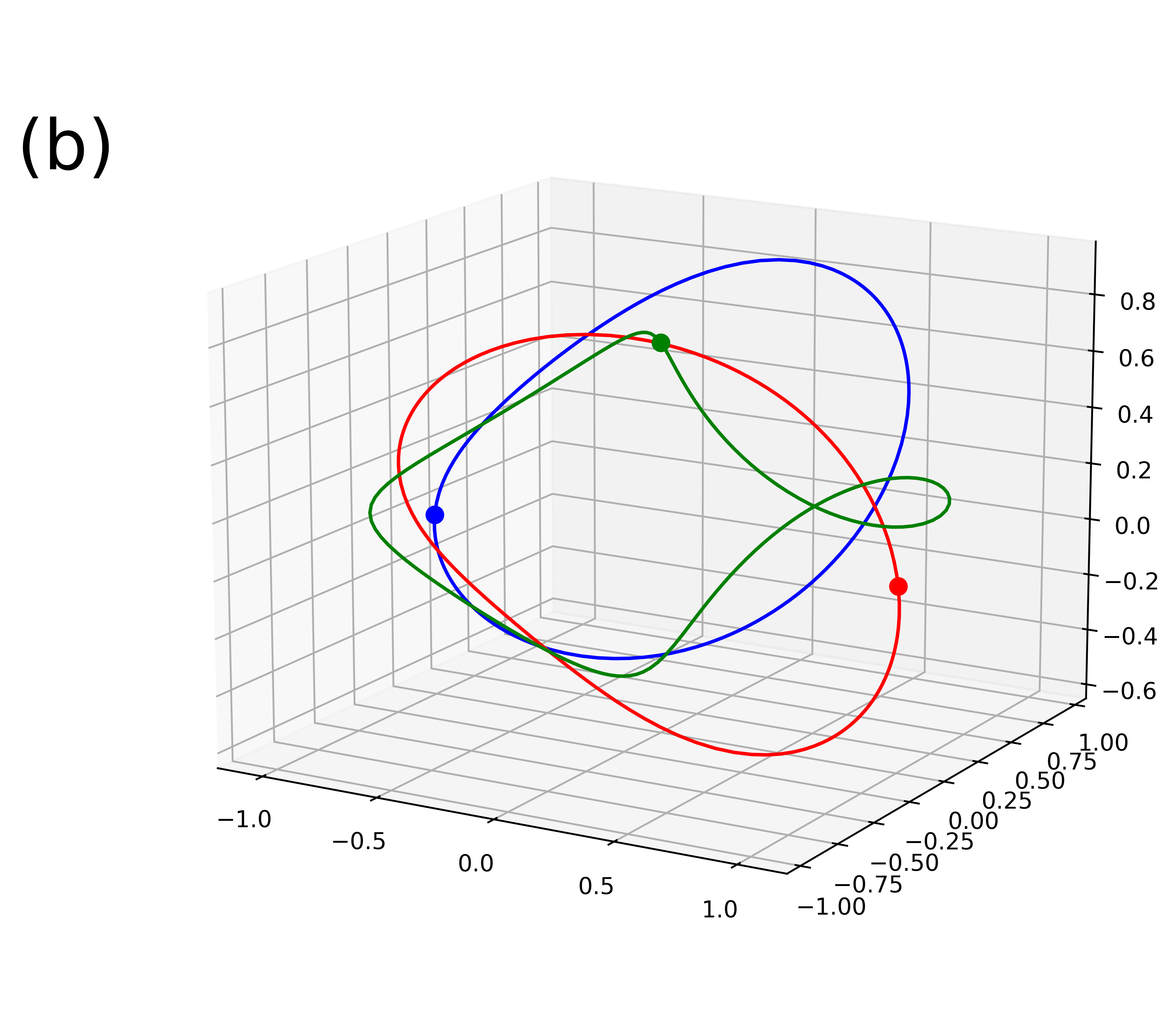}
  \includegraphics[width=5cm]{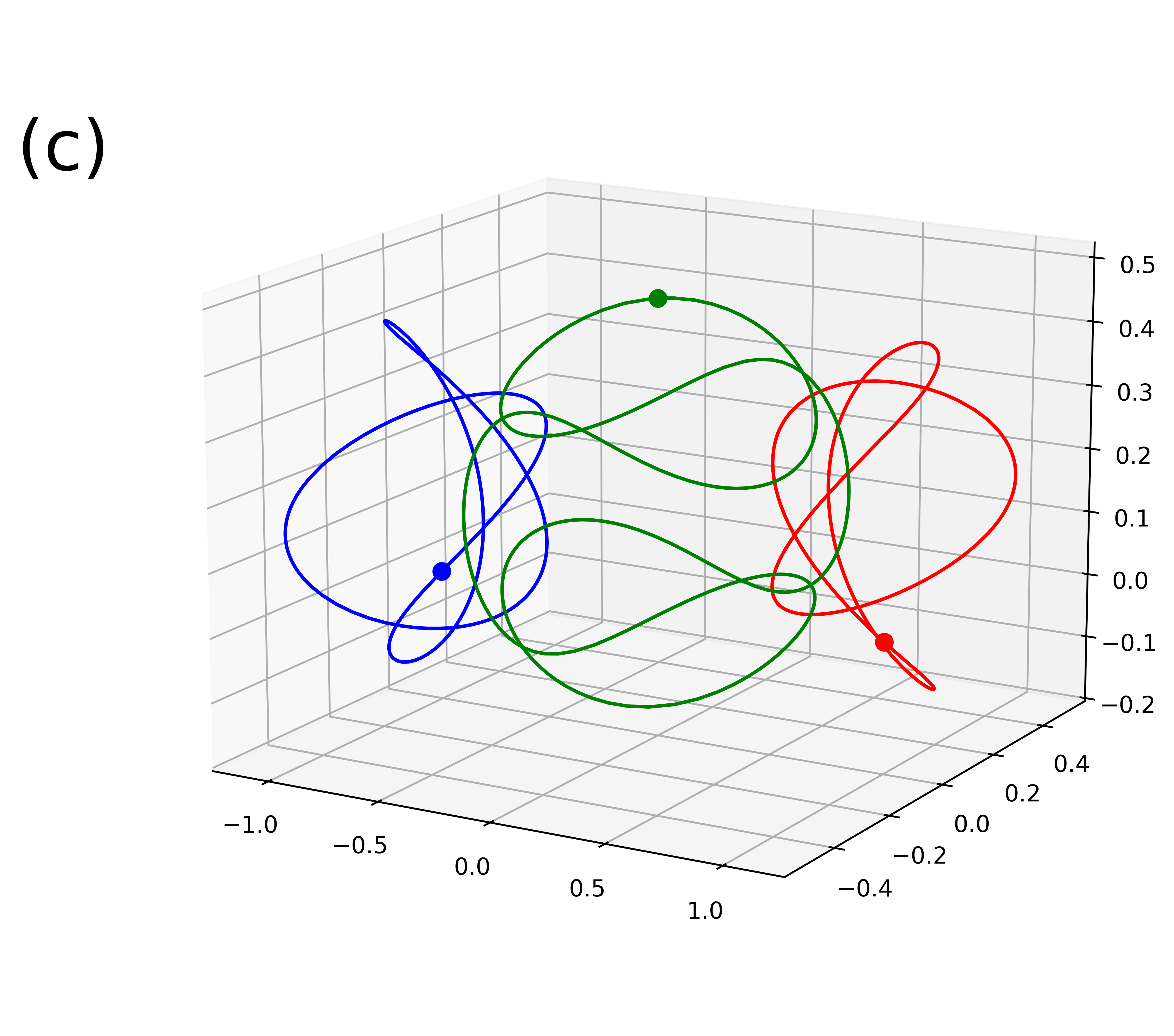}\\
  \includegraphics[width=5cm]{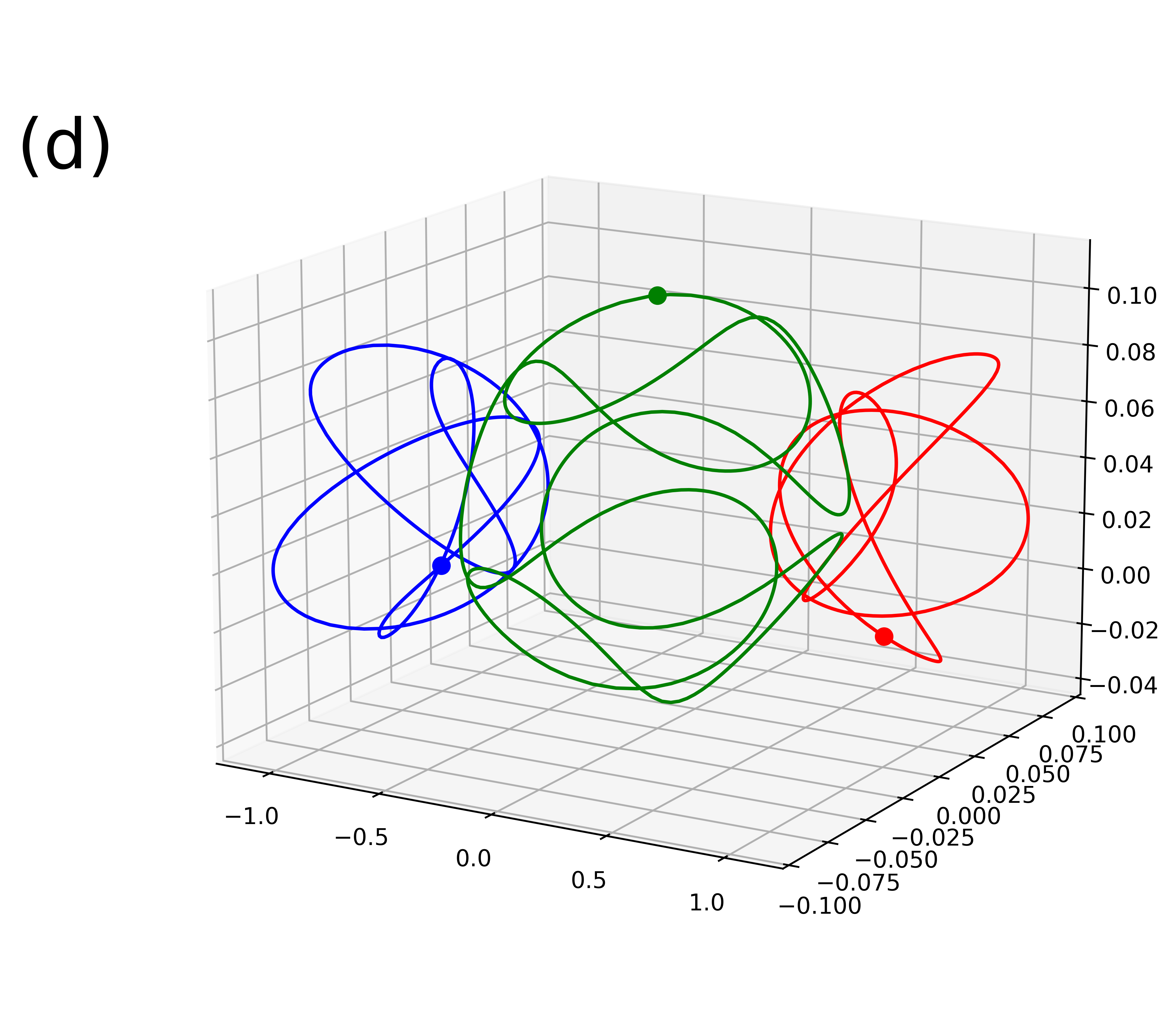}
  \includegraphics[width=5cm]{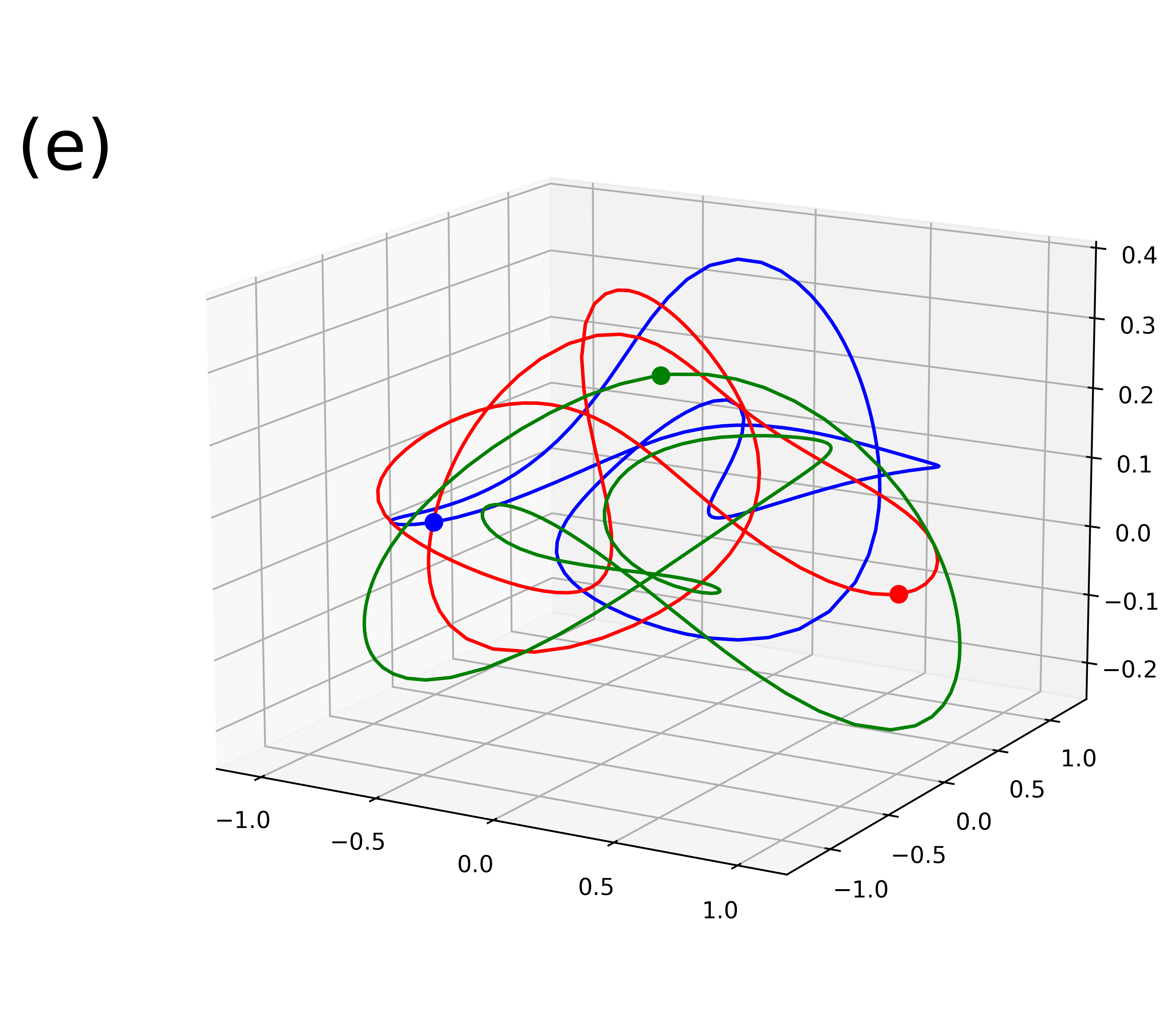}
  \includegraphics[width=5cm]{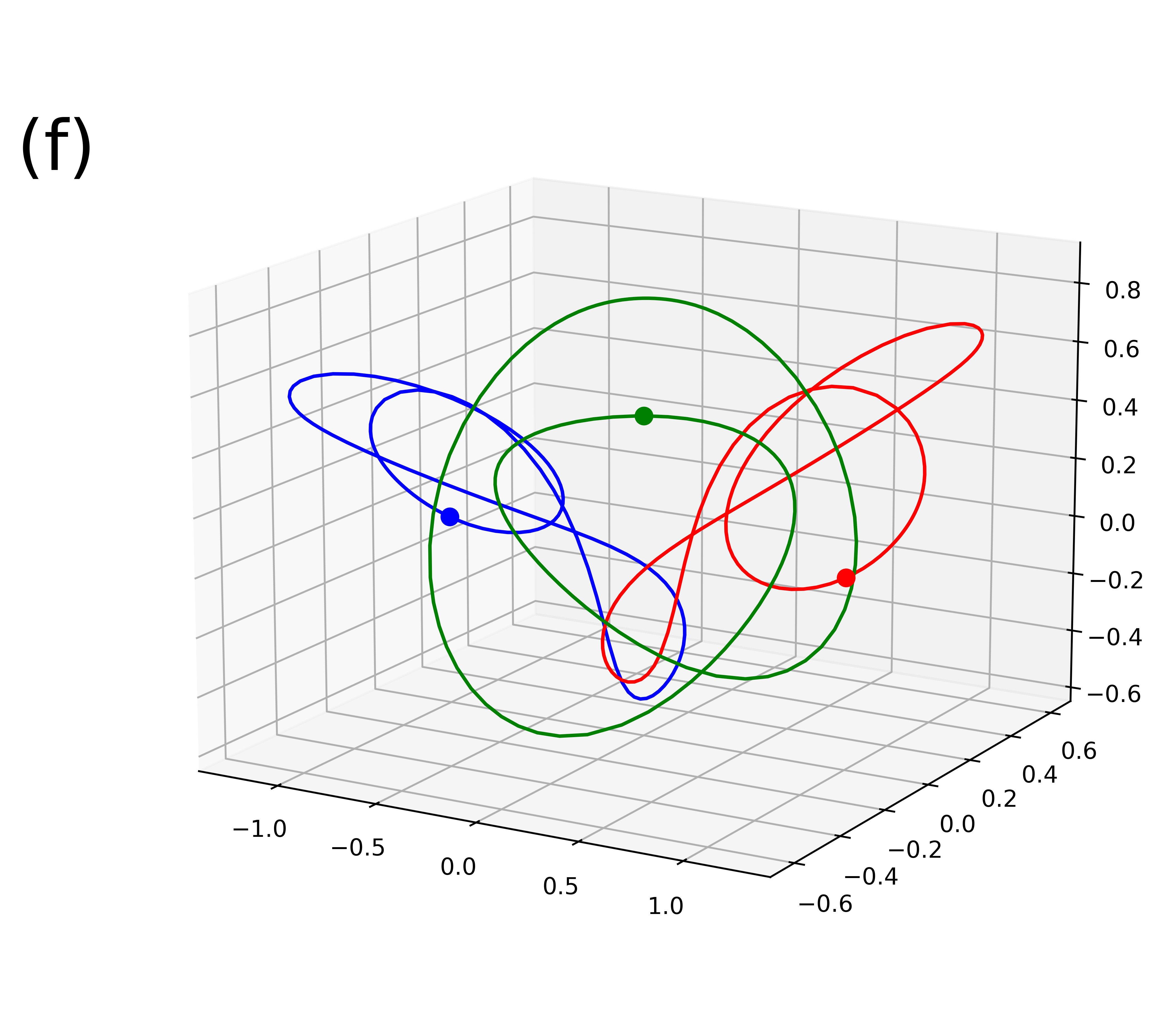}
  \caption{3D periodic orbits of general three-body problem: (a) $O_{2}(1.2)$, (b) $O_{8}(0.6)$, (c) $O_{3}(1.0)$, (d) $O_{4}(1.0)$, (e) $O_{6}(1.0)$, (f) $O_{6}(1.2)$.  Blue: Body-1; Red: Body-2; Green: Body-3.  Filled circle: initial positions of the three bodies. }
  \label{fig:orbits}
\end{figure}

Note that $m_{3}$ is a variable although $m_{1}=m_{2}=1$ is fixed.   Without loss of generality, we consider the case  $m_3=0.1 n$ where $1\leq n\leq 20$ is an integer, respectively, thus totally $2\times 10^{10}$ cases are done for the the grid search method.  Using a national supercomputer,  we  discovered 10,059 three-dimensional periodic orbits of the general three-body system.  
 It is found that 1,996 among them are linearly stable, approximately 20\% of the total.
Their initial condition,  period $T$ and stability are given on the website  \textcolor{blue}{https://github.com/sjtu-liao/three-body}.   These 3D periodic orbits are named $O_{n}(m_3)$, where $n$ corresponds to the $n$-th orbit when ordered by its value of period $T$ (from small to large)  for a fixed mass $m_3$ (when $m_{1}=m_{2}=1$).  For example, $O_1(1)$ denotes the discovered 3D periodic orbit  with the shortest period $T$ when $m_3=1$.   
Some 3D periodic orbits are shown in Fig.~\ref{fig:orbits}  and their corresponding initial conditions $(z_0, v_x, v_y, v_z)$ and periods $T$ are given in Table~\ref{general-ICs} in Supplementary Information.  To the best of our knowledge, all of these 3D periodic orbits  have  {\em never} been reported and thus are novel.  It should be emphasized  that we can choose an {\em arbitrary} value of $m_{3}$, such as $m_{3}=0.615$ or $m_{3}=0.872$, to obtain 3D periodic orbits in the similarly way.  So, in theory, we can gain an {\em arbitrarily} large amount of   3D periodic orbits of three-body system using the same initial configuration as that mentioned in this paper.

\begin{figure}[tb]
\centering
   \includegraphics[width=5cm]{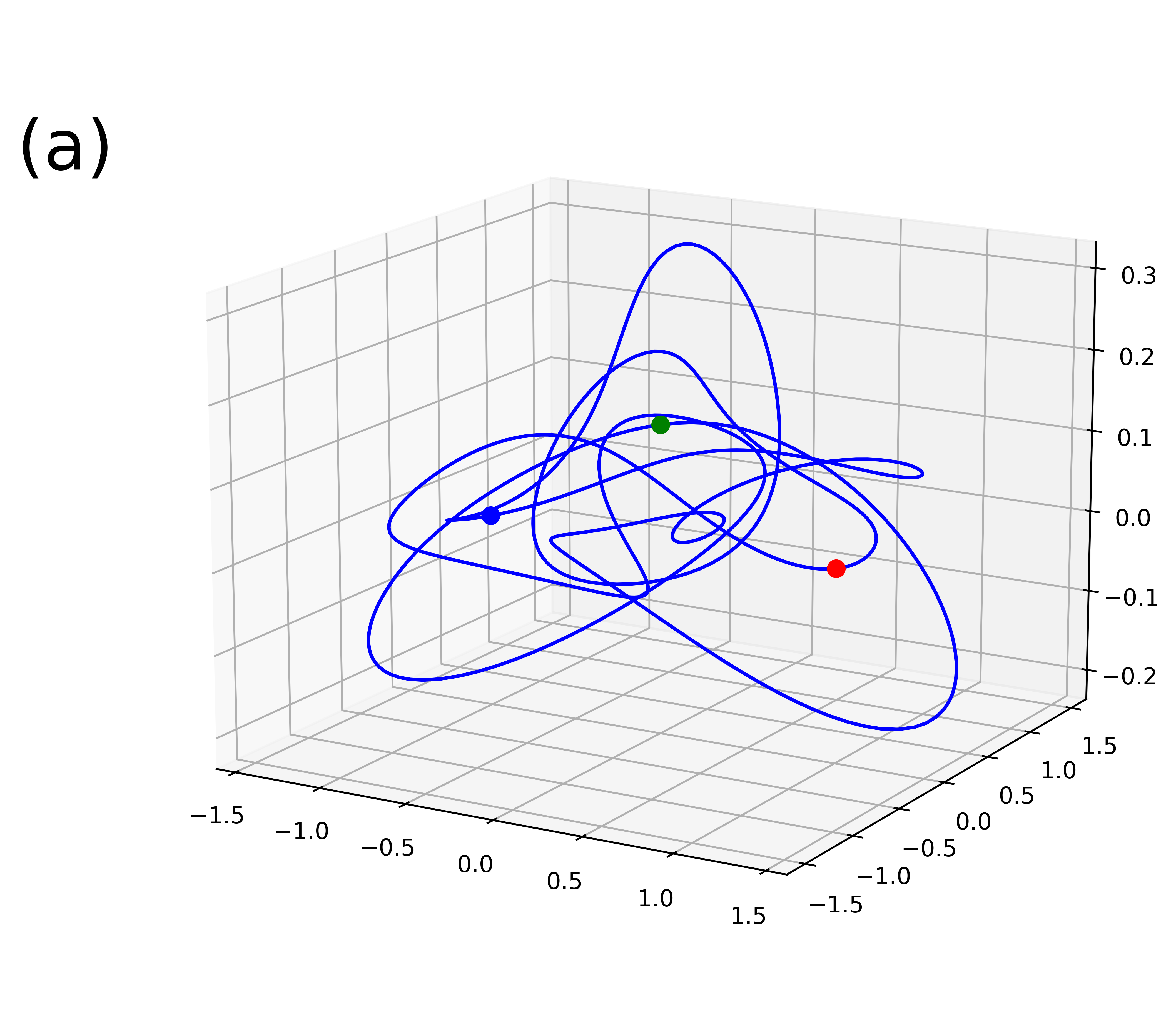}
   \includegraphics[width=5cm]{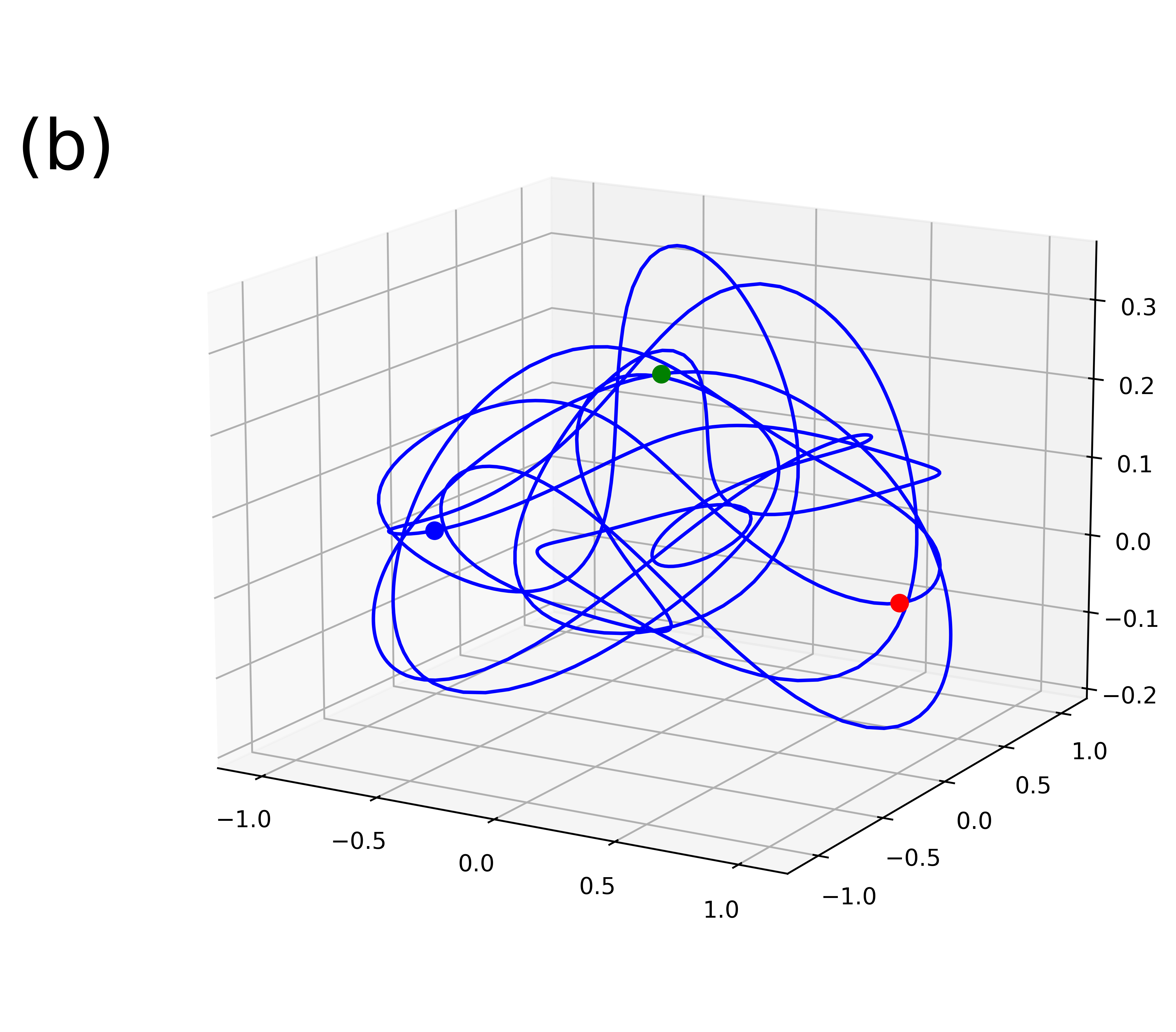}\\
   \includegraphics[width=5cm]{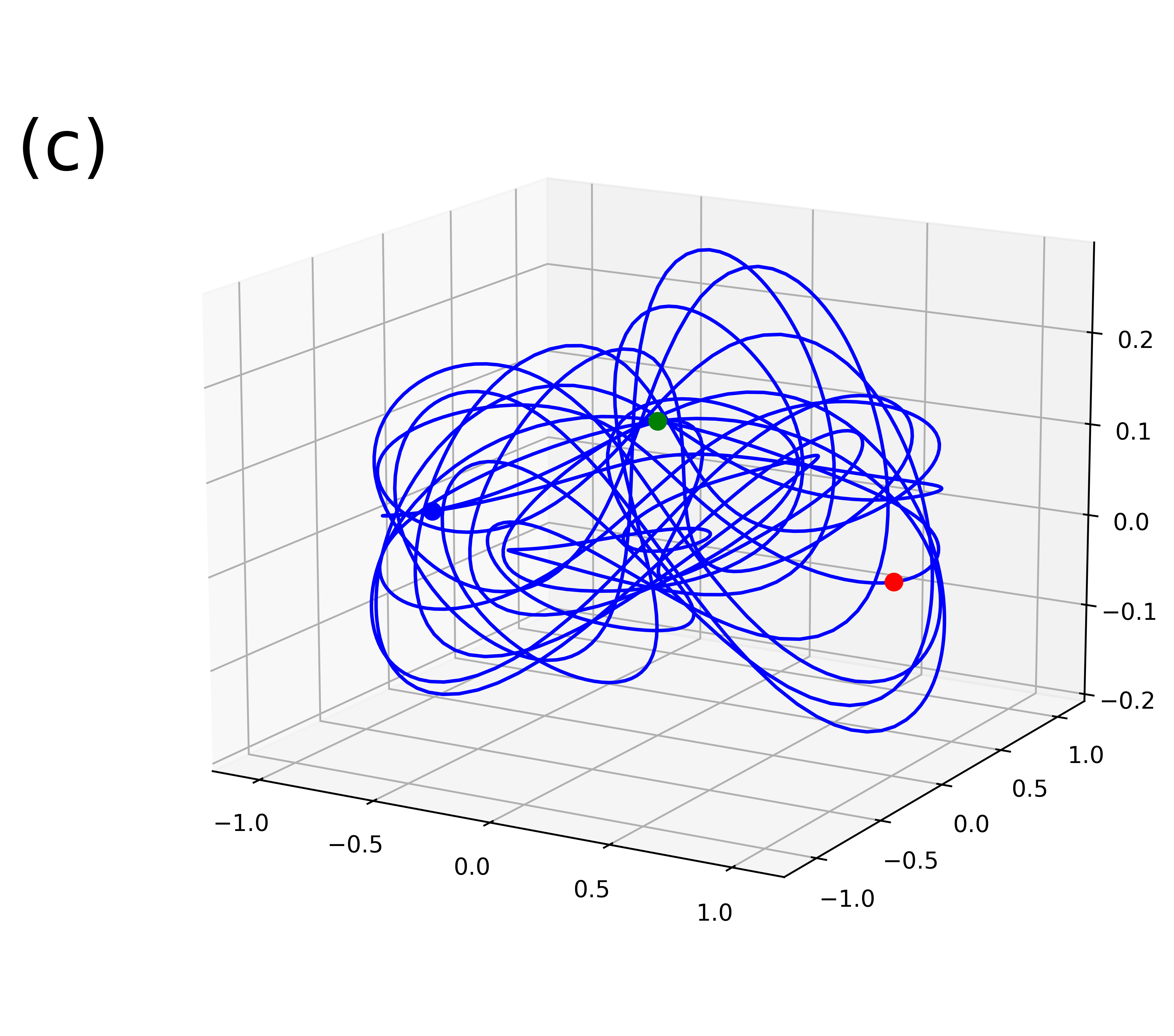}
   \includegraphics[width=5cm]{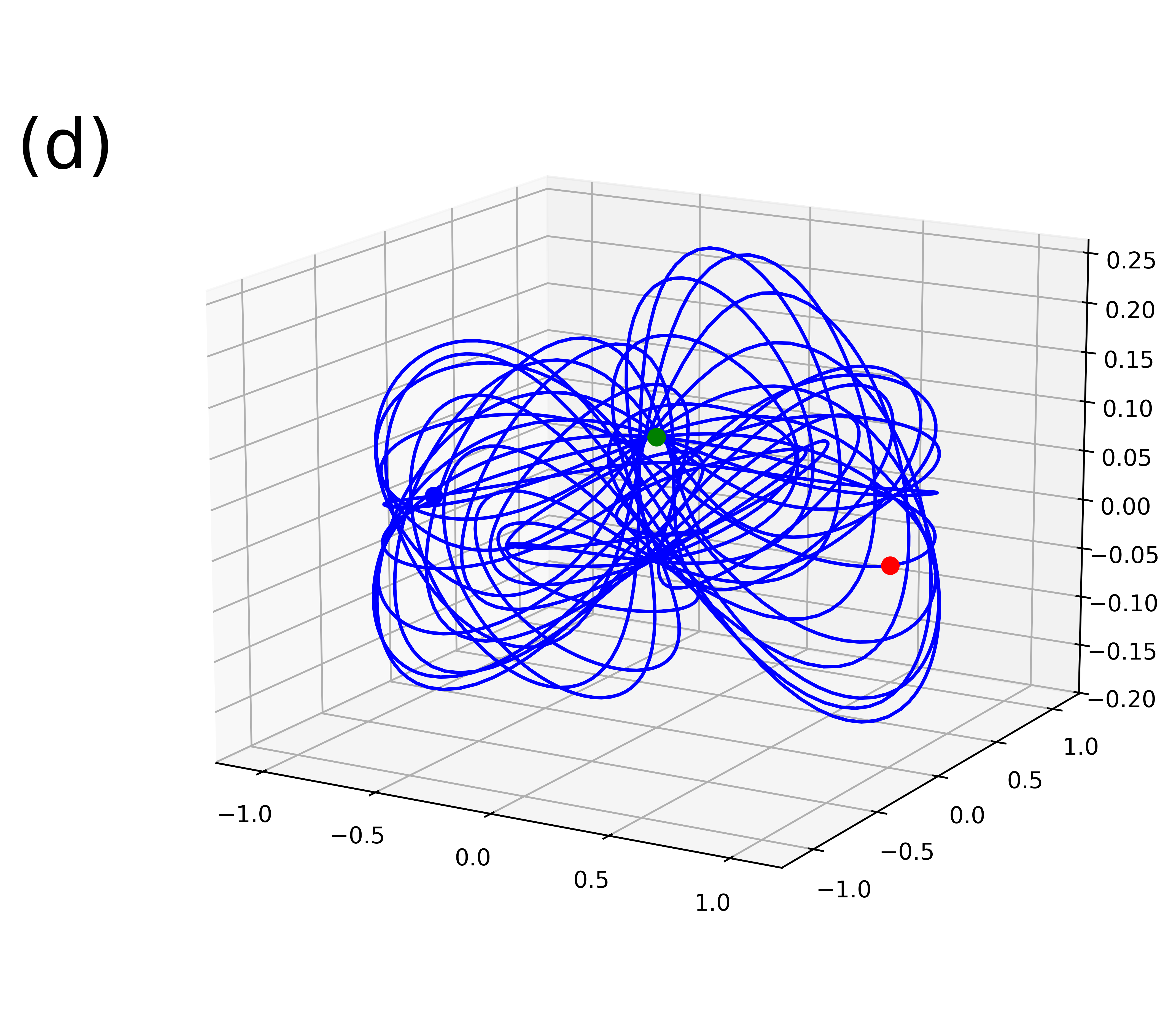}
  \caption{3D choreographic periodic orbits of three-body problem  in case of $m_{1}=m_{2}=m_{3}=1$: (a) $O_{62}(1.0)$, (b) $O_{64}(1.0)$, (c) $O_{231}(1.0)$, (d) $O_{524}(1.0)$. Three bodies move along a single closed orbit (blue).  Filled circles:  the initial positions of the three bodies.  Blue: Body-1; Red: Body-2; Green: Body-3.  }\label{fig:choe}
\end{figure}

\begin{figure}[htb]
\centering
   \includegraphics[width=5cm]{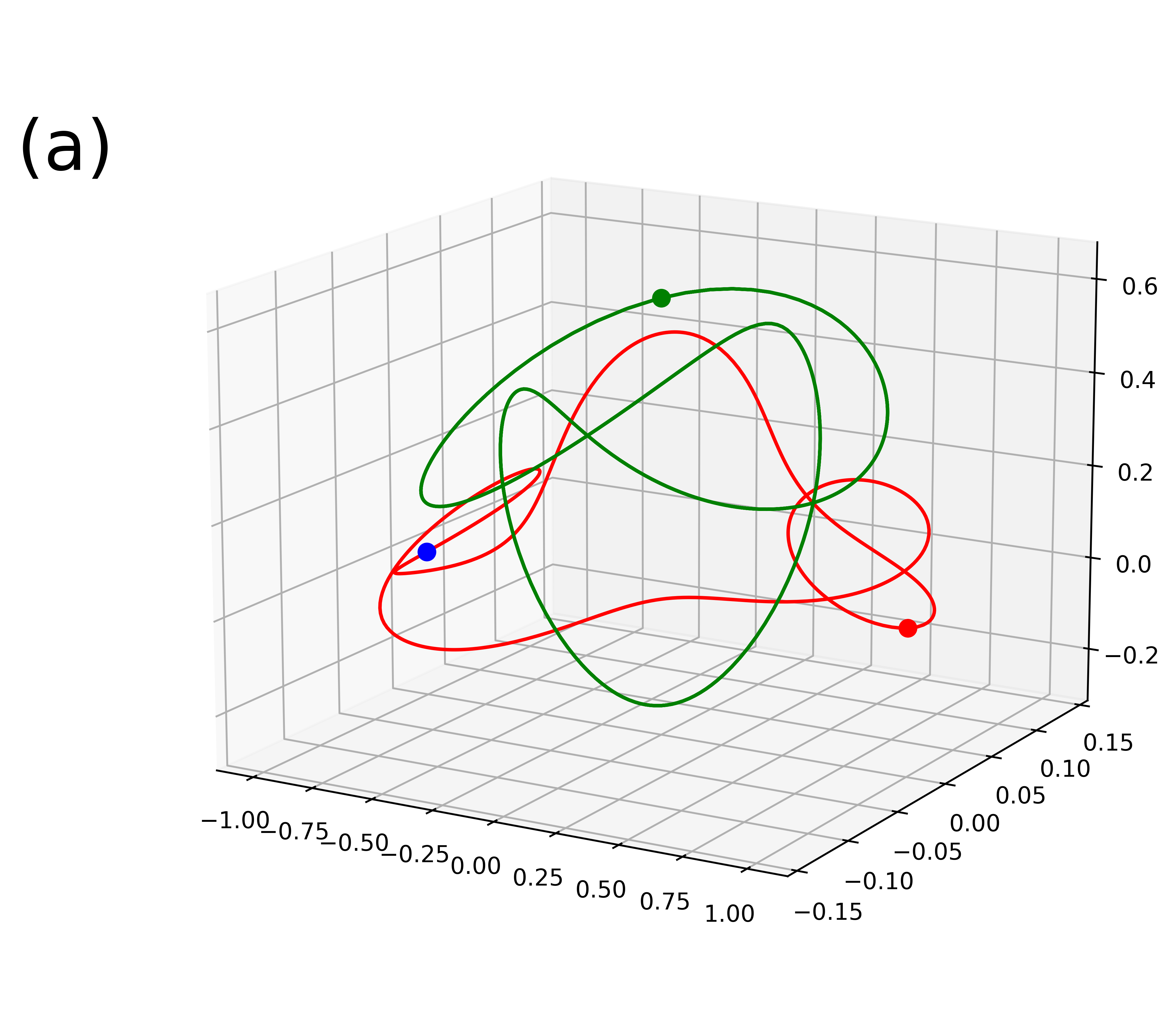}
   \includegraphics[width=5cm]{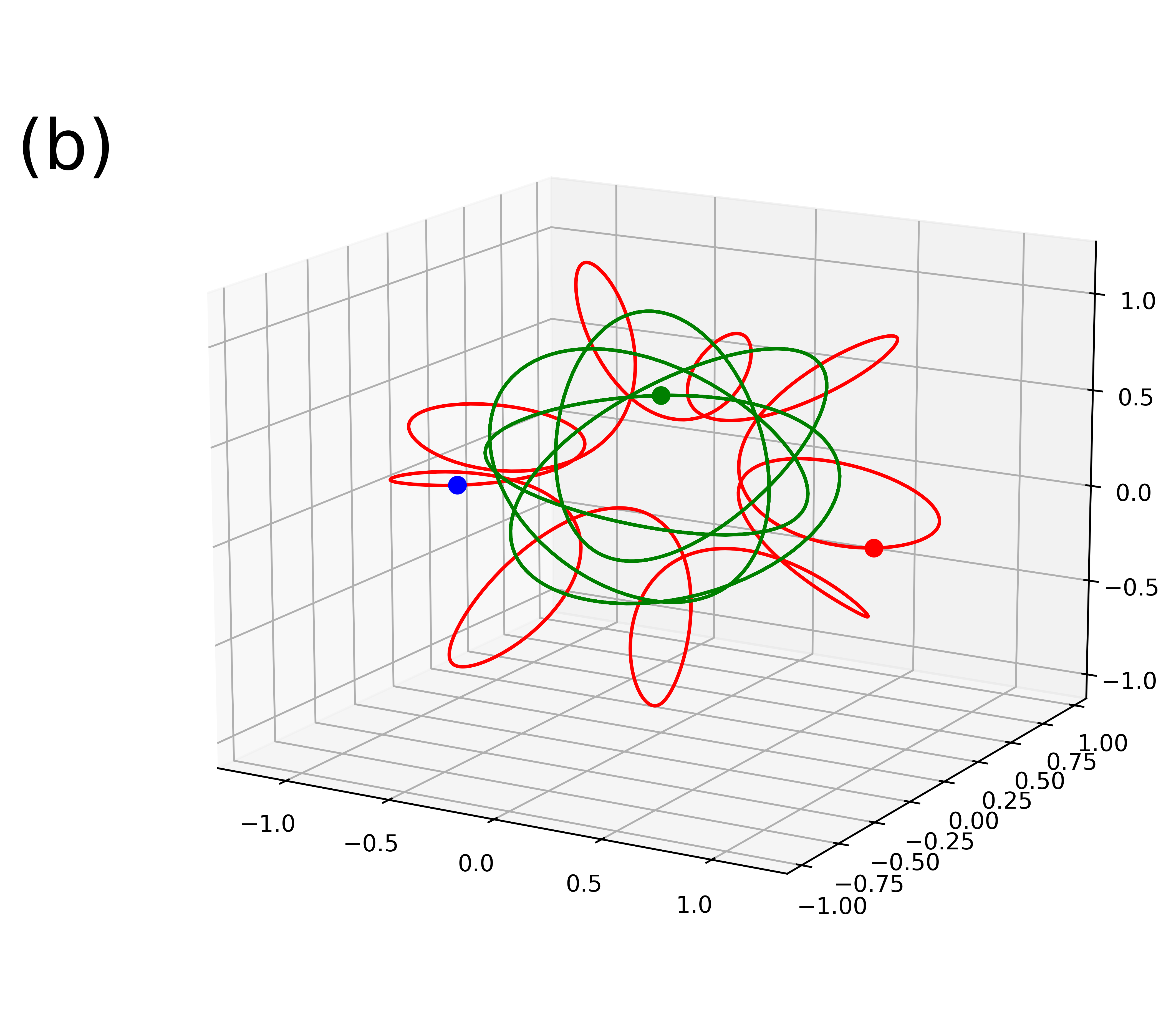}\\
   \includegraphics[width=5cm]{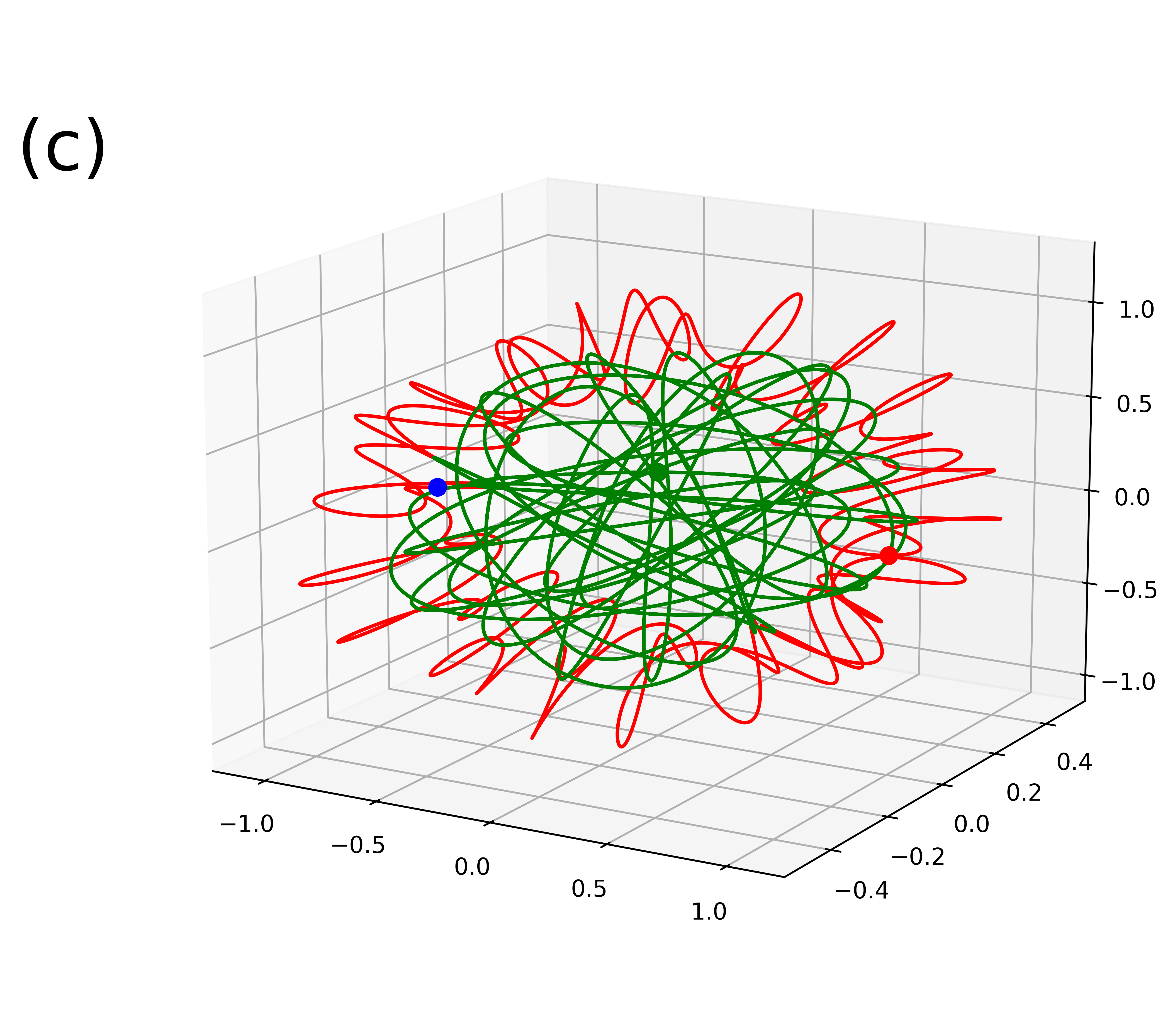}
   \includegraphics[width=5cm]{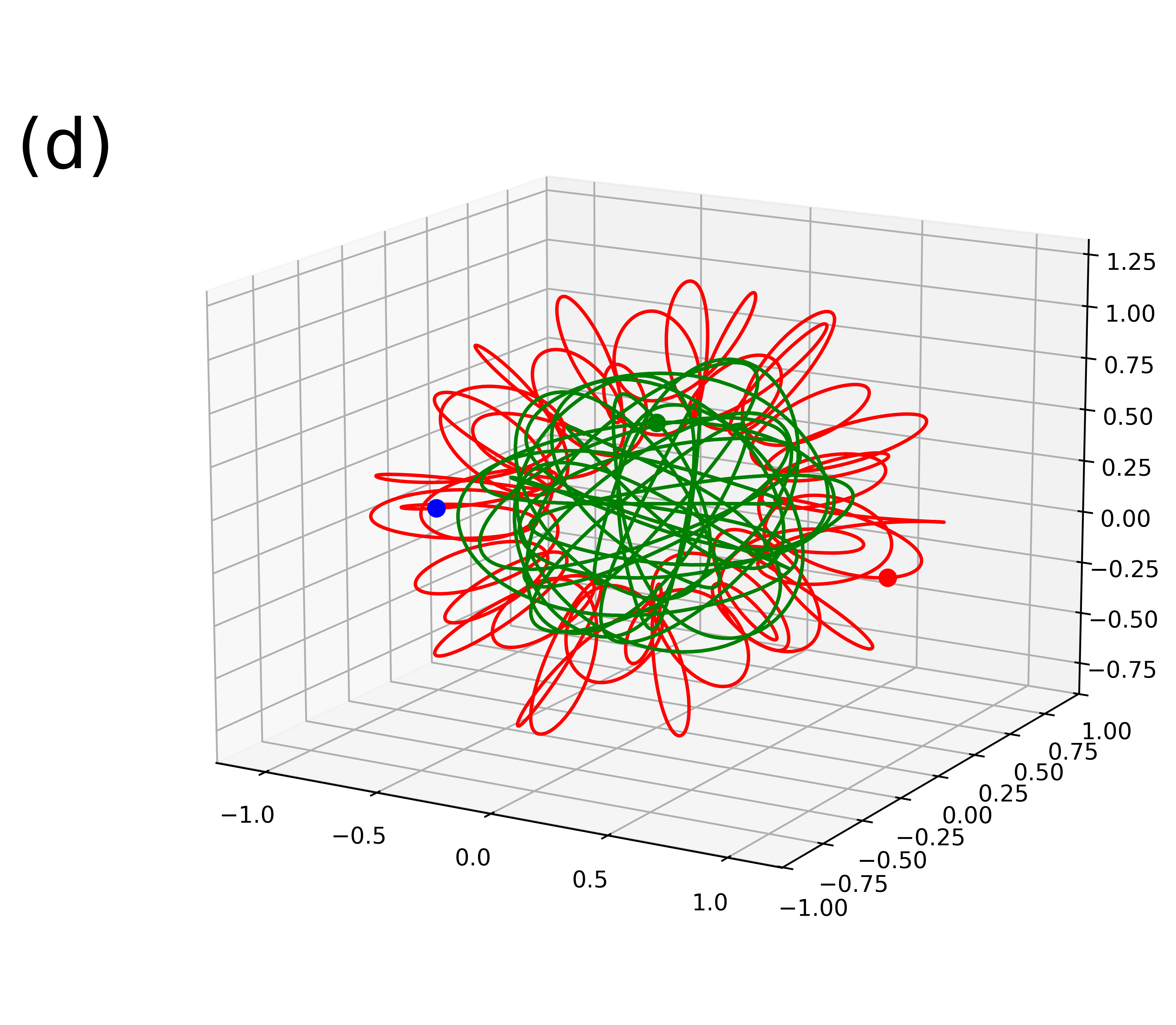}
  \caption{3D ``piano-trio orbits''  of three-body problem with  two bodies ($m_{1}=m_{2}=1$) moving along a single closed orbit (red) but the third ($m_{3}\neq 1$) along a distinct orbit (green): (a) $O_{6}(0.6)$, (b) $O_{26}(1.1)$, (c) $O_{48}(0.5)$, (d) $O_{267}(0.9)$. Filled circles:  the initial positions of the three bodies. Blue: Body-1 ($m_{1}=1$) ; Red: Body-2 ($m_{2}=1$) ; Green: Body-3 ($m_{3}\neq 1$)). 
}\label{fig:binary}
\end{figure}

In the case of three equal masses, i.e. $m_{1}=m_{2}=m_{3}=1$,  we discover 1,504 three-dimensional periodic orbits.  Among them, there are 21  ``choreographic''  orbits, say, the three bodies move periodically in a single closed orbit. Their initial conditions, periods and stability are given in Table~\ref{choe-ICs} in Supplementary Information.  Different from the famous planar choreographic orbit Figure-8   \cite{Moore1993, Chenciner2000}, these twenty-one choreographic orbits are {\em three-dimensional},  as shown in Fig.~\ref{fig:choe}.    To the best of our knowledge, such kind of 3D choreographic periodic orbits of the general three-body system have {\em never} been reported.

Besides, it is very interesting that, when $m_{1}=m_{2}=1$ and $m_3 \neq 1$, we found 273 three-dimensional periodic orbits in the 19 cases of $m_{3}=0.1 n$ ($1\leq n \leq 20$ but $n\neq 10$ is an integer),  where the two bodies with equal mass ($m_1=m_2=1$) move along a single closed orbit but the third ($m_{3}\neq 1$) moves along a different closed orbit.  We name them ``piano-trio orbit'', like a trio for two violins and one piano.   Their initial conditions,  periods and stability are given on the website  \textcolor{blue}{https://github.com/sjtu-liao/three-body}.   Four orbits of them are as shown in Fig.~\ref{fig:binary}, where Body-1 ($m_1=1$) and Body-2 ($m_2=1$) follow a single closed orbit (in red) but Body-3 ($m_3 \neq 1$) traces a distinct path (in green),  and their initial conditions, periods and stability are given in Table~\ref{binary-ICs} in Supplementary Information.   To the best of our knowledge, such kind of 3D   periodic  orbits  of general three-body system have {\em never} been reported.

\section{Concluding remarks and discussion}

In this paper,  by means of a numerical strategy based on  the grid search method, Newton-Raphson method \cite{Abad2011}  and clean numerical method (CNS) \cite{Liao2009, Liao2014-SciChina, Li2018clean, Hu2020, qin2020influence, Liao2023}, we successfully discovered 10,059 three-dimensional periodic orbits of the three-body problem in the cases of $m_{1}=m_{2}=1$ and $m_{3}=0.1n$  ($1\leq n\leq 20$), among which  1,996 (about 20\%)  are linearly stable.  It should be emphasized that our numerical approach is valid for {\em arbitrary} mass  $m_{3}$ (such as  $m_{3}=0.615$ or $m_{3}=0.872$) so that in theory we can gain an {\em arbitrarily} large amount of 3D periodic orbits of the three-body problem in a similar way (we will do it and update the dataset on the website  \textcolor{blue}{https://github.com/sjtu-liao/three-body}).  Certainly, these 3D periodic orbits should be helpful for us to have better understandings about the chaotic properties of the famous three-body problems, since they ``are, so to say, the only opening through which we can try to penetrate in a place which, up to now, was supposed to be inaccessible'', as pointed out by Poincar{\'e}, the founder of chaos theory.

It should be emphasized that, in the case of three equal masses ($m_{1}=m_{2}=m_{3}=1$),  we  discovered twenty-one 3D ``choerographical''  periodic orbits whose three bodies move periodically in a single closed orbit.  To the best of our knowledge, these 3D ``choerographical''  periodic orbits have {\em never} been reported, which reveals its general existence for three-body problem and indicates the novelty of this work.   Besides,  in the case of two equal masses ($m_{1}=m_{2}=1$ and $m_{3}\neq 1$), we discovered  273  three-dimensional periodic orbits with two bodies ($m_{1}=m_{2}=1$) moving along a single closed orbit and the third $m_{3} \neq 1$ along a different one: we name them ``piano-trio'' orbits, like a trio for two violins and one piano.  To the best of our knowledge, these ``piano-trio'' orbits have {\em never} been reported,  indicating the novelty of this work.  

As illustrated by Liao et al. \cite{Liao2022}, one can gain arbitrarily accurate planar periodic orbit of three-body problem by means of CNS.  Similarly, each of our discovered 10,059 three-dimensional periodic orbits reported in this paper could be in an arbitrary accuracy.  For example,  the high accuracy (in 70 significant digits) initial condition and the period $T$ of the  linearly stable periodic orbit $O_{3}(1.0)$ are listed in Table~\ref{high-precision-IC} in Supplementary Information. In physics,  the minimum spatial distance is a Planck length $l_{p} = 1.616252 \times 10^{-35}$ meter and the maximum distance is the diameter  of the observable  universe (as the characteristic length)  $d_{u}\approx 9.3 \times10^{10}$ light year  $ \approx 8.4 \times 10^{26}$ meter, so that the minimum dimensionless spatial length that has  physical meaning  is  $l_{p}/d_{u} \approx 1.9 \times 10^{-62}$.  Note that our 3D periodic orbit $O_{3}(1.0)$ listed in Table~\ref{high-precision-IC} in Supplementary Information is accurate in the 70 significant digits, which thus can be regarded, from the physical viewpoint,  as accurate as a closed-form solution.         

Many further investigations should be done in future.  It is true that in this paper we only consider the case with two equal masses $m_{1}=m_{2}=1$.  However,  using the known 3D periodic orbits obtained in this paper as a starting point,  it is straightforward to gain new 3D periodic orbits of general three-body problem with three {\em unequal} masses by means of the continuation method \cite{Li2021}.   Note that, for planar periodic orbits of three-body problem, their topological identification can be classified by braid groups \cite{Moore1993} or shape-space sphere \cite{Montgomery1998, Suvakov2013}.   However,  for a 3D periodic orbit of general three-body problem, it is  an open question how to identify its topological classification \cite{Moore1993}.  

\section*{Acknowledgment}
The calculations were performed on the Tianhe-2 Supercomputer, National  Supercomputer Center in Guangzhou, China.  This work is partly supported by National Natural Science Foundation of China (No. 12002132) and State Key Laboratory of Ocean Engineering.



\bibliographystyle{elsarticle-num}

\bibliography{ref}

\begin{thebibliography}{10}
\expandafter\ifx\csname url\endcsname\relax
  \def\url#1{\texttt{#1}}\fi
\expandafter\ifx\csname urlprefix\endcsname\relax\def\urlprefix{URL }\fi
\expandafter\ifx\csname href\endcsname\relax
  \def\href#1#2{#2} \def\path#1{#1}\fi

\bibitem{Newton1687}
I.~Newton, Philosophi{\ae} naturalis principia mathematica (Mathematical
  principles of natural philosophy), London: Royal Society Press, 1687.

\bibitem{Euler1767}
L.~Euler, De motu rectilineo trium corporum se mutuo attrahentium, Novi
  commentarii academiae scientiarum Petropolitanae (1767) 144--151.

\bibitem{Lagrange1772}
J.-L. Lagrange, Essai sur le probleme des trois corps, Prix de l’acad{\'e}mie
  royale des Sciences de paris 9 (1772) 292.

\bibitem{Poincare1890}
J.~H. Poincar{\' e}, {Sur} le probl{\' e}me des trois corps et les {\'
  e}quations de la dynamique. {D}ivergence des s{\' e}ries de {M. Lindstedt},
  Acta Math. 13 (1890) 1--270.

\bibitem{Broucke1973}
R.~Broucke, H.~Lass, A note on relative motion in the general three-body
  problem, Celestial mechanics 8~(1) (1973) 5--10.

\bibitem{Hadjidemetriou1975}
J.~D. Hadjidemetriou, T.~Christides, Families of periodic orbits in the planar
  three-body problem, Celestial mechanics 12~(2) (1975) 175--187.

\bibitem{Henon1976}
M.~H{\'e}non, A family of periodic solutions of the planar three-body problem,
  and their stability, Celestial mechanics 13~(3) (1976) 267--285.

\bibitem{Moore1993}
C.~Moore, Braids in classical dynamics, Phys. Rev. Lett. 70 (1993) 3675--3679.

\bibitem{Chenciner2000}
A.~Chenciner, R.~Montgomery, A remarkable periodic solution of the three-body
  problem in the case of equal masses, Annals of Mathematics 152~(3) (2000)
  881--901.

\bibitem{Suvakov2013}
M.~\ifmmode~\check{S}\else \v{S}\fi{}uvakov, V.~Dmitra\ifmmode \check{s}\else
  \v{s}\fi{}inovi\ifmmode~\acute{c}\else \'{c}\fi{}, Three classes of newtonian
  three-body planar periodic orbits, Phys. Rev. Lett. 110 (2013) 114301.

\bibitem{Iasko2014}
P.~P. Iasko, V.~V. Orlov,
  \href{https://doi.org/10.1134/S1063772914110080}{Search for periodic orbits
  in the general three-body problem}, Astronomy Reports 58~(11) (2014)
  869--879.
\newblock \href {https://doi.org/10.1134/S1063772914110080}
  {\path{doi:10.1134/S1063772914110080}}.
\newline\urlprefix\url{https://doi.org/10.1134/S1063772914110080}

\bibitem{Hudomal2015}
A.~Hudomal, New periodic solutions to the three-body problem and gravitational
  waves, {M.S.} {T}hesis, University of Belgrade, Serbia (2015).

\bibitem{Suvakov2016}
M.~{\v{S}}uvakov, M.~Shibayama, Three topologically nontrivial choreographic
  motions of three bodies, Celestial Mechanics and Dynamical Astronomy 124~(2)
  (2016) 155--162.

\bibitem{Rose2016}
D.~Rose, Geometric phase and periodic orbits of the equal-mass, planar
  three-body problem with vanishing angular momentum, {Ph.D.} {T}hesis,
  University of Sydney, Australia (2016).

\bibitem{Lorenz2006Tellus}
E.~N. Lorenz, Computational periodicity as observed in a simple system, Tellus
  A: Dynamic Meteorology and Oceanography 58A (2006) 549 -- 557.

\bibitem{Liao2009}
S.~Liao, On the reliability of computed chaotic solutions of non-linear
  differential equations, Tellus A 61~(4) (2009) 550--564.

\bibitem{Liao2014-SciChina}
S.~Liao, P.~Wang, On the mathematically reliable long-term simulation of
  chaotic solutions of lorenz equation in the interval [0,10000], Sci. China -
  Phys. Mech. Astron. 57 (2014) 330 -- 335,.

\bibitem{Li2018clean}
X.~Li, S.~Liao, Clean numerical simulation: a new strategy to obtain reliable
  solutions of chaotic dynamic systems, Applied Mathematics and Mechanics
  39~(11) (2018) 1529--1546.

\bibitem{Hu2020}
T.~Hu, S.~Liao, On the risks of using double precision in numerical simulations
  of spatio-temporal chaos, Journal of Computational Physics 418 (2020) 109629.

\bibitem{qin2020influence}
S.~Qin, S.~Liao, Influence of numerical noises on computer-generated simulation
  of spatio-temporal chaos, Chaos Solitons Fractals 136 (2020) 109790.

\bibitem{Liao2023}
S.~Liao, Clean Numerical Simulation, CRC Press, 2023.

\bibitem{Li2017}
X.~Li, S.~Liao, More than six hundred new families of newtonian periodic planar
  collisionless three-body orbits, SCIENCE CHINA Physics, Mechanics \&
  Astronomy 60~(12) (2017) 129511.

\bibitem{Li2018}
X.~Li, Y.~Jing, S.~Liao, Over a thousand new periodic orbits of a planar
  three-body system with unequal masses, Publications of the Astronomical
  Society of Japan 70~(4) (2018) 64.

\bibitem{Li2021}
X.~Li, X.~Li, S.~Liao, One family of 13315 stable periodic orbits of
  non-hierarchical unequal-mass triple systems, Science China Physics,
  Mechanics \& Astronomy 64~(1) (2021) 219511.

\bibitem{Liao2022}
S.~Liao, X.~Li, Y.~Yang, Three-body problem: From newton to supercomputer plus
  machine learning, New Astronomy 96 (2022) 101850.

\bibitem{Hristov2024a}
I.~Hristov, R.~Hristova, V.~Dmitra{\v{s}}inovi{\'c}, K.~Tanikawa, Three-body
  periodic collisionless equal-mass free-fall orbits revisited, Celestial
  Mechanics and Dynamical Astronomy 136~(1) (2024) 7.

\bibitem{Sitnikov1961}
K.~{Sitnikov}, {The Existence of Oscillatory Motions in the Three-Body
  Problem}, Soviet Physics Doklady 5 (1961) 647.

\bibitem{Corbera2004}
M.~Corbera, J.~Llibre,
  \href{http://dx.doi.org/10.1137/S0036141002407880}{Families of periodic
  orbits for the spatial isosceles 3-body problem}, SIAM Journal on
  Mathematical Analysis 35~(5) (2004) 1311–1346.
\newblock \href {https://doi.org/10.1137/s0036141002407880}
  {\path{doi:10.1137/s0036141002407880}}.
\newline\urlprefix\url{http://dx.doi.org/10.1137/S0036141002407880}

\bibitem{Yan2015}
D.~Yan, R.~Liu, X.~Hu, W.~Mao, T.~Ouyang, New phenomena in the spatial
  isosceles three-body problem with unequal masses, International Journal of
  Bifurcation and Chaos 25~(12) (2015) 1550169.

\bibitem{Perdomo2017}
O.~M. Perdomo, \href{http://dx.doi.org/10.1007/s12346-017-0244-1}{A bifurcation
  in the family of periodic orbits for the spatial isosceles 3 body problem},
  Qualitative Theory of Dynamical Systems 17~(2) (2017) 411–428.
\newblock \href {https://doi.org/10.1007/s12346-017-0244-1}
  {\path{doi:10.1007/s12346-017-0244-1}}.
\newline\urlprefix\url{http://dx.doi.org/10.1007/s12346-017-0244-1}

\bibitem{Katopodis1979}
K.~Katopodis, \href{http://dx.doi.org/10.1007/BF01230173}{Continuation of
  periodic orbits: Three-dimensional circular restricted to the general
  three-body problem}, Celestial Mechanics 19~(1) (1979) 43–51.
\newblock \href {https://doi.org/10.1007/bf01230173}
  {\path{doi:10.1007/bf01230173}}.
\newline\urlprefix\url{http://dx.doi.org/10.1007/BF01230173}

\bibitem{Markellos1980}
V.~V. Markellos, \href{http://dx.doi.org/10.1007/BF01230227}{The
  three-dimensional general three-body problem: Determination of periodic
  orbits}, Celestial Mechanics 21~(3) (1980) 291–309.
\newblock \href {https://doi.org/10.1007/bf01230227}
  {\path{doi:10.1007/bf01230227}}.
\newline\urlprefix\url{http://dx.doi.org/10.1007/BF01230227}

\bibitem{Hairer1993}
E.~Hairer, G.~Wanner, S.~P. Norsett, Solving Ordinary Differential Equations I:
  Non-stiff Problems, Springer-Verlag Berlin Heidelberg, 1993.

\bibitem{Abad2011}
A.~Abad, R.~Barrio, A.~Dena, Computing periodic orbits with arbitrary
  precision, Phys. Rev. E 84 (2011) 016701.

\bibitem{Guckenheimer1983}
J.~Guckenheimer, P.~Holmes,
  \href{http://dx.doi.org/10.1007/978-1-4612-1140-2}{Nonlinear Oscillations,
  Dynamical Systems, and Bifurcations of Vector Fields}, Springer New York,
  1983.
\newblock \href {https://doi.org/10.1007/978-1-4612-1140-2}
  {\path{doi:10.1007/978-1-4612-1140-2}}.
\newline\urlprefix\url{http://dx.doi.org/10.1007/978-1-4612-1140-2}

\bibitem{Richter1993}
K.~Richter, G.~Tanner, D.~Wintgen, Classical mechanics of two-electron atoms,
  Physical Review A 48~(6) (1993) 4182.

\bibitem{Montgomery1998}
R.~Montgomery, The n-body problem, the braid group, and action-minimizing
  periodic solutions, Nonlinearity 11~(2) (1998) 363.

\end{thebibliography}

\newpage

\begin{center}

{\LARGE Supplementary Information}

\end{center}

\renewcommand\thetable{S\arabic{table}}
\setcounter{table}{0}

\renewcommand\theequation{S\arabic{equation}}
\setcounter{equation}{0}

\begin{table}[htb]
\tabcolsep 0pt \caption{
The initial conditions, the periods $T$ and the stability of some 3D periodic orbits in the case of the initial position $\bm{r}_1 = (-1, 0, 0)$, $\bm{r}_2=(1, 0, 0)$ and $\bm{r}_3 = (0, 0, z_0)$ with the initial velocities $\bm{v}_1 = (v_x, v_y, v_z)$, $\bm{v}_2 = (v_x, v_y, -v_z)$ and $\bm{v}_3 = (-2 v_x/m_3, -2 v_y/m_3, 0)$.  Stability of periodic orbit: S (linearly stable), U (linearly unstable).} \label{general-ICs} \vspace*{-12pt}
\begin{center}
\def\temptablewidth{1\textwidth}
{\rule{\temptablewidth}{1pt}}
\begin{tabular*}{\temptablewidth}{@{\extracolsep{\fill}}ccccccc}
Orbit  $O_{n}(m_{3})$  & $z_0$ & $v_x$ & $v_y$ & $v_z$ & $T$ & stability \\
\hline
$O_{2}(1.2)$ & 1.0220057827E+00 & -2.7260000746E-01 & -4.3209371195E-01 & 6.2947340717E-01 & 6.9057763983E+00 & U \\
$O_{8}(0.6)$ & 7.2379454056E-01 & 2.1745309714E-01 & -2.4023509732E-01 & 4.8072135301E-01 & 8.2513531298E+00 & U \\
$O_{3}(1.0)$ & 4.7687826428E-01 & 4.0213691007E-01 & 1.8035695129E-01 & 2.1044512814E-01 & 6.8316220363E+00 & S \\
$O_{4}(1.0)$ & 1.0656465072E-01 & 4.1396073100E-01 & 4.7046903637E-02 & 4.4299118421E-02 & 7.2303054580E+00 & S \\
$O_{6}(1.0)$ & 2.5738110869E-01 & 3.0241724398E-01 & 5.6056948094E-01 & 5.2896109217E-03 & 1.3654015842E+01 & U \\
$O_{6}(1.2)$ & 4.4254976548E-01 & 6.3300867799E-01 & 1.7485764338E-01 & -2.6069495427E-01 & 8.2359759431E+00 & U \\
\end{tabular*}
{\rule{\temptablewidth}{1pt}}
\end{center}
\end{table}

\begin{table}[htb]
\tabcolsep 0pt \caption{
The initial conditions, the periods $T$ and the stability of 21 choreographic 3D periodic orbits having three bodies moving along the same path in the case of the initial position   $\bm{r}_1 = (-1, 0, 0)$, $\bm{r}_2=(1, 0, 0)$ and $\bm{r}_3 = (0, 0, z_0)$ with the initial velocities $\bm{v}_1 = (v_x, v_y, v_z)$, $\bm{v}_2 = (v_x, v_y, -v_z)$ and $\bm{v}_3 = (-2 v_x/m_3, -2 v_y/m_3, 0)$.  Stability of periodic orbit: S (linearly stable), U (linearly unstable).} \label{choe-ICs} \vspace*{-12pt}
\begin{center}
\def\temptablewidth{1\textwidth}
{\rule{\temptablewidth}{1pt}}
\begin{tabular*}{\temptablewidth}{@{\extracolsep{\fill}}ccccccc}
Orbit  $O_{n}(m_{3})$  & $z_0$ & $v_x$ & $v_y$ & $v_z$ & $T$ & stability \\
\hline
$O_{62}(1.0)$ & 1.4254560021E-01 & 3.4973952580E-01 & 6.0245969580E-01 & 5.6258558122E-03 & 4.4413258151E+01 & U \\
$O_{64}(1.0)$ & 2.4089819317E-01 & 2.9736744570E-01 & 5.5127998288E-01 & 6.4720700746E-03 & 4.5160913797E+01 & U \\
$O_{231}(1.0)$ & 1.3449345804E-01 & 3.3746407711E-01 & 5.3495391505E-01 & 1.7586983000E-03 & 8.2638396171E+01 & S \\
$O_{264}(1.0)$ & 1.2527262314E-01 & 3.3812559500E-01 & 5.3478083136E-01 & 1.8426277977E-03 & 8.8847804260E+01 & S \\
$O_{468}(1.0)$ & 1.9882559981E-01 & 3.5789268529E-01 & 5.3413197456E-01 & -5.3389716378E-04 & 1.1603279244E+02 & U \\
$O_{524}(1.0)$ & 9.2921053426E-02 & 3.4199333600E-01 & 5.3399733168E-01 & 1.6826415762E-03 & 1.2041512314E+02 & S \\
$O_{574}(1.0)$ & 2.4895906866E-01 & 3.3871794052E-01 & 5.5917249111E-01 & -2.5336443257E-03 & 1.2535891398E+02 & S \\
$O_{609}(1.0)$ & 7.9082023338E-02 & 2.8653274947E-01 & 5.3295320774E-01 & 1.3330149482E-03 & 1.2885423891E+02 & U \\
$O_{617}(1.0)$ & 2.5059548392E-01 & 3.1883837350E-01 & 5.5145840743E-01 & 7.2285919945E-04 & 1.2927664287E+02 & U \\
$O_{623}(1.0)$ & 2.4884379653E-01 & 2.9277777970E-01 & 5.5413846299E-01 & 6.9945423951E-03 & 1.2975160287E+02 & U \\
$O_{735}(1.0)$ & 1.4523478543E-01 & 3.3263557496E-01 & 5.2616069788E-01 & -3.1374383107E-03 & 1.4041031143E+02 & U \\
$O_{793}(1.0)$ & 1.5066856881E-01 & 3.3509372155E-01 & 5.3528992956E-01 & 1.2102376067E-03 & 1.4629315855E+02 & U \\
$O_{941}(1.0)$ & 1.3957142176E-01 & 3.3672076170E-01 & 5.3505271584E-01 & 1.6586446900E-03 & 1.5893652668E+02 & S \\
$O_{1034}(1.0)$ & 7.9952100998E-02 & 3.9991383731E-01 & 5.2892786472E-01 & 2.6785166522E-03 & 1.6684804313E+02 & U \\
$O_{1062}(1.0)$ & 2.5387857578E-01 & 3.0855862740E-01 & 5.5506896654E-01 & 3.4527925512E-03 & 1.6887732550E+02 & S \\
$O_{1114}(1.0)$ & 1.9719159455E-01 & 3.4208132539E-01 & 5.3782184023E-01 & 1.4597658877E-03 & 1.7279868625E+02 & S \\
$O_{1172}(1.0)$ & 6.3233041014E-02 & 3.4473818853E-01 & 5.3330581391E-01 & 1.2154707447E-03 & 1.7726783502E+02 & S \\
$O_{1265}(1.0)$ & 6.1061814942E-02 & 3.4489838389E-01 & 5.3326623797E-01 & 1.1774765759E-03 & 1.8358838190E+02 & S \\
$O_{1414}(1.0)$ & 4.9792913732E-02 & 2.6873050476E-01 & 5.2603515916E-01 & 3.2150647009E-04 & 1.9557410885E+02 & S \\
$O_{1488}(1.0)$ & 1.5788544119E-01 & 3.3561402713E-01 & 5.3228206363E-01 & -2.2004697254E-03 & 2.0197023210E+02 & S \\
$O_{1497}(1.0)$ & 1.1019540397E-01 & 3.3940414208E-01 & 5.3491778263E-01 & 2.0614879729E-03 & 2.0302355527E+02 & S \\
\end{tabular*}
{\rule{\temptablewidth}{1pt}}
\end{center}
\end{table}

\begin{table}[htb]
\tabcolsep 0pt \caption{
The initial conditions, the periods $T$ and the stability  of some 3D periodic orbits having two bodies moving along the same path in the case of the initial position $\bm{r}_1 = (-1, 0, 0)$, $\bm{r}_2=(1, 0, 0)$ and $\bm{r}_3 = (0, 0, z_0)$ with the initial velocities  $\bm{v}_1 = (v_x, v_y, v_z)$, $\bm{v}_2 = (v_x, v_y, -v_z)$ and $\bm{v}_3 = (-2 v_x/m_3, -2 v_y/m_3, 0)$. Stability of periodic orbits: S (linearly stable), U (linearly unstable).} \label{binary-ICs} \vspace*{-12pt}
\begin{center}
\def\temptablewidth{1\textwidth}
{\rule{\temptablewidth}{1pt}}
\begin{tabular*}{\temptablewidth}{@{\extracolsep{\fill}}ccccccc}
Orbit  $O_{n}(m_{3})$ & $z_0$ & $v_x$ & $v_y$ & $v_z$ & $T$ & stability \\
\hline
$O_{6}(0.6)$ & 6.1460435884E-01 & 1.0381048232E-01 & 7.2494633230E-02 & 5.8966905104E-02 & 7.9743488400E+00 &   U \\
$O_{26}(1.1)$ & 6.1717279647E-01 & 5.0563910986E-01 & 2.8923907397E-01 & 6.3478431464E-03 & 2.7381502412E+01 &  U \\
$O_{48}(0.5)$ & 2.5356269746E-01 & 3.0249330624E-01 & 7.6131511629E-02 & -3.9679960521E-03 & 9.0118410122E+01 & S \\
$O_{267}(0.9)$ & 5.7288742602E-01 & 3.2057036195E-01 & 2.3346447921E-01 & 3.1450924543E-03 & 8.6966037898E+01 & S \\
\end{tabular*}
{\rule{\temptablewidth}{1pt}}
\end{center}

\tabcolsep 0pt \caption{
The high accuracy (in 70 significant digits) initial condition and the period $T$ of the  linearly stable periodic orbit $O_{3}(1.0)$ in the case of the initial position $\bm{r}_1 = (-1, 0, 0)$, $\bm{r}_2=(1, 0, 0)$ and $\bm{r}_3 = (0, 0, z_0)$ with the initial velocities $\bm{v}_1 = (v_x, v_y, v_z)$, $\bm{v}_2 = (v_x, v_y, -v_z)$, $\bm{v}_3 = (-2 v_x/m_3, -2 v_y/m_3, 0)$ and  $m_{3}=1.0$.} \label{high-precision-IC} \vspace*{-12pt}
\begin{center}
\def\temptablewidth{1\textwidth}
{\rule{\temptablewidth}{1pt}}
\begin{tabular*}{\temptablewidth}{@{\extracolsep{\fill}}cl}
Orbit  $O_{3}(1.0)$ & \\
\hline
$z_0$ & 4.768782642803115460616110933545915887541644925960260369253641303299294E-1
\\
$v_x$ & 4.021369100747237493832253390929016257978052115111599603699588130212840E-1
\\
$v_y$ & 1.803569512862586641317199478891917026060004921964179954624832790547685E-1
\\
$v_z$ & 2.104451281378731221497174690764795110575771159319962594676582115390083E-1\\
$T$ & 6.831622036284445040766025062746064085737914286107653347931784678966507
\\
\end{tabular*}
{\rule{\temptablewidth}{1pt}}
\end{center}
\end{table}

\end{document}